\documentclass[12pt]{amsart}
\usepackage{amsmath,amssymb,eucal}

\newcommand{\HH}{\mathcal{H}}

\begin{document}

\title[3+1 decomposition]{3+1 decomposition in the new action\\ for the Einstein Theory of Gravitation}
\author{L.~D.~Faddeev}
\address{St.Petersburg Department of Steklov Mathematical Institute}

\begin{abstract}
    The action of recently proposed formulation of Einstein Theory of
    Gravitation is written according to 3+1 decomposition of the space-time
    variables. The result coincides with known formulation of Dirac and
    Arnowitt-Deser-Misner.
\end{abstract}

\maketitle

    Recently I proposed to use new dynamical variables to describe
    the gravitational field
\cite{heeq},
\cite{nvars}.
    In
\cite{nvars}
    the new formulation was shown to be equivalent to the classical one
    of Hilbert-Einstein. So the question arises why to do such an effort.
    The only answer I can give now is that I follow an old advice of
    Feynman --- to generalize a theory one must work it out in many guises.
    And there is no doubt that we need to develop Einstein's theory
    further.

    In this note I continue the work in 
\cite{heeq},
\cite{nvars}
    and develop 3+1 decomposition of the space-time variables in the action
    functional. I shall show how traditional formulas of Dirac
\cite{Dirac}
    and Arnowitt-Deser-Misner
\cite{ADM}
    appear in my formulation.

    The set of dynamical variables introduced in
\cite{heeq},
\cite{nvars}
    consists of 40 components --- 10 covariant vector fields
$ f_{\mu}^{A}(x) $
    on four dimensional space-time 
$ M_{4} $
    with coordinates
$ x^{\mu} $; 
    thus
$ A=1,\ldots , 10 $,
$ \mu = 0,1,2,3 $.

    In terms of these variables I define metric
\begin{equation*}
    g_{\mu\nu} = f_{\mu}^{A}f_{\nu}^{A}
\end{equation*}
    and linear connection
\begin{equation*}
    \Omega_{\alpha\mu}^{\beta} = f^{\beta A} \partial_{\mu} f_{\alpha}^{A} .
\end{equation*}
    Here
$ f^{\mu A}(x) $
    are contravariant vector fields
\begin{equation*}
    f^{\mu A} = g^{\mu\nu} f_{\nu}^{A} ,
\end{equation*}
    where
$ g^{\mu\nu} $ is, as usual, inverse to
$ g_{\mu\nu} $
\begin{equation*}
    g_{\mu\sigma}g^{\sigma\nu} = \delta_{\mu}^{\nu} .
\end{equation*}
    As in
\cite{heeq},
\cite{nvars}
    I do not bother with the subtleties of the pseudoriemannian signature,
    so all scalar products are euclidean. In such situation the separation of
    time and space variables
$ x^{0} $ and
$ x^{k} $
    seems somewhat artificial, but I continue to follow this convention to
    avoid minus signs. So I use a term ``3+1 decomposition'' instead
    of the space-time one.

    The action is written more transparently via the contravariant components
\begin{equation*}
    A = \int \sqrt{g} S d^{4}x ,
\end{equation*}
    where
\begin{equation*}
    S = \Pi^{AB} (\partial_{\mu} f^{\mu A}\partial_{\nu}f^{\nu B}
	- \partial_{\mu} f^{\nu A}\partial_{\nu}f^{\mu B})
\end{equation*}
    and
$ \Pi^{AB} $ is the ``vertical'' projector
\begin{equation*}
    \Pi^{AB} = \delta^{AB} - g_{\mu\nu} f^{\mu A} f^{\nu B} .
\end{equation*}
    The expression
$ S $
    defines the scalar curvature of the connection
$ \Omega_{\alpha\mu}^{\beta} $.
    The full curvature tensor
\begin{equation*}
    S_{\alpha\beta,\mu\nu} = g_{\beta\gamma} \bigl(
	\partial_{\mu} \Omega_{\alpha\nu}^{\gamma}
	-\partial_{\nu} \Omega_{\alpha\mu}^{\gamma}
    + \Omega_{\alpha\mu}^{\sigma} \Omega_{\sigma\nu}^{\gamma}
    - \Omega_{\alpha\nu}^{\sigma} \Omega_{\sigma\mu}^{\gamma}
    \bigr)
\end{equation*}
    is beautiffully expressed as
\begin{equation*}
    S_{\alpha\beta,\mu\nu} = \Pi^{AB} (
	\partial_{\mu} f_{\alpha}^{A} \partial_{\nu} f_{\beta}^{B}
	-\partial_{\mu} f_{\beta}^{A} \partial_{\nu} f_{\alpha}^{B}
    )
\end{equation*}
    and
\begin{equation*}
    S = g^{\mu\alpha}g^{\nu\beta} S_{\alpha\beta,\mu\nu} .
\end{equation*}
    All these formulas include only usual partial derivatives, but they are
    fully covariant with respect to the general coordinate transformations
\begin{equation*}
    \delta f_{\mu}^{A} = - \partial_{\mu}\epsilon^{\nu} f_{\nu}^{A}
	- \epsilon^{\nu} \partial_{\nu} f_{\mu}^{A} ,
\end{equation*}
    where
$ \epsilon^{\nu} $
    is a vector field, defining infinitesimal coordinate transformation.

    In this note I shall explicitly realize the 3+1 decomposition of
    these formulas in coordinates 
$ x^{\mu} = (x^{0},x^{k}), k=1,2,3 $
    and refer to
$ x^{0}=t $
    as time and to
$ x^{k} $
    as space variables.
    The main goal is to rewrite the action in hamiltonian-like form.

    The first observation is that 
$ S $
    contains time derivatives only linearly.
    This allows to develop the reduction formalism following the general ideas
    of my paper with R.~Jackiw
\cite{FJ}.
    There the dynamical variables entering the original lagrangian are divided
    into three classes: canonical, excludable and Lagrange multipliers.
    To exclude the variables of second class one is allowed to use equations
    of motion which do not contain the time derivatives.

    Among the equations of motion, which are derived in
\cite{nvars},
    there is a set of equations which express the vanishing of the torsion
    of the connection
$ \Omega_{\alpha\mu}^{\beta} $
\begin{equation*}
    \Omega_{\alpha\mu}^{\beta} = \Omega_{\mu\alpha}^{\beta} .
\end{equation*}
    Out of these 24 equations 12 do not contain the time derivatives
\begin{equation*}
    \Omega_{ik}^{0} = \Omega_{ki}^{0} , \quad
	\Omega_{ik}^{l} = \Omega_{ki}^{l}
\end{equation*}
    and I shall use them in the reduction of action in what follows.

    The formulas I plan to derive should be covariant with respect to
    coordinate transformation, generated by vector fields
$ \epsilon^{i}(x) $,
    obtained from
$ \epsilon^{\mu}(x) $
    by restriction
\begin{equation*}
    \epsilon^{0}(x) = 0 , \quad \partial_{t} \epsilon^{i}(x) = 0 .
\end{equation*}
    The covariant vector fields
$ f_{k}^{A} $
    are compatible with this requirement
\begin{equation*}
    \delta f_{k}^{A} = -\partial_{k} \epsilon^{l}(x) f_{l}^{A}
	- \epsilon^{l} \partial_{l} f_{k}^{A} .
\end{equation*}
    Furthermore the component
$ f^{0A} $
    defines scalars
\begin{equation*}
    \delta f^{0A} = -\epsilon^{k} \partial_{k} f^{0A} .
\end{equation*}
    
    I begin by writing
\begin{equation*}
    \sqrt{g} S = \Sigma + \HH ,
\end{equation*}
    where
$ \Sigma $
    contains all terms with time derivatives
\begin{equation*}
    \Sigma = 2 \sqrt{g} \Pi^{AB} g^{\mu l} (
	\partial_{l}f_{\mu}^{A} \partial_{t} f^{0B} 
	- \partial_{l}f^{0A} \partial_{t} f_{\mu}^{B} 
    ) 
\end{equation*}
    and
\begin{equation*}
    \HH =  \sqrt{g} g^{k\sigma}g^{l\rho} (
	\partial_{k}f_{\sigma}^{A} \partial_{l} f_{\rho}^{B} 
	- \partial_{l}f_{\sigma}^{A} \partial_{k} f_{\rho}^{B} 
    ) .
\end{equation*}
    We can interpret
$ \Sigma $
    as one form using substitution
\begin{equation*}
    \partial_{t} f_{\mu}^{A} \to d f_{\mu}^{A} .
\end{equation*}
    In this guise the action
$ A $
    is an explicit example of general scheme in
\cite{FJ}.

    Now we proceed to realize the promised separation.
    The covariant 3-dimensional metric
$ \gamma_{ik} $
    is given by
\begin{equation*}
    \gamma_{ik} = f_{i}^{A} f_{k}^{A}
\end{equation*}
    and the components of 4-dimensional contravariant metric
$ g^{\mu k} $,
    which we need, can be expressed via
$ \gamma^{ik} $, $ g^{0i} $, $ g^{00} $, which are 3-minesional
    tensor, vector and scalar, correspondingly,
\begin{equation*}
    g^{ik} = \gamma^{ik} + \frac{g^{0i}g^{0k}}{g^{00}} .
\end{equation*}
    The 4-dimensional determinant can be written as
\begin{equation*}
    g = \gamma / g^{00} ,
\end{equation*}
    where
$ \gamma $
    is determinant of metric
$ \gamma_{ik} $.
    We shall also see, that terms containing
$ f_{0}^{A} $
    will always have the form
\begin{equation*}
    g^{00}f_{0}^{A} + g^{0k}f_{k}^{A} = f^{0A} .
\end{equation*}
    Let us begin our rearrangement with one-form
$ \Sigma $.
    We have
\begin{multline*}
    \Sigma = 2\sqrt{g} \Pi^{AB} \bigl[ (\gamma^{ml}+\frac{g^{0m}g^{0l}}{g^{00}})
	(\partial_{l}f_{m}^{A}\partial_{t}f^{0B}
	- \partial_{l}f^{0A}\partial_{t}f_{m}^{B}) \\
	  + g^{0l} (\partial_{l}f_{0}^{A}\partial_{t}f^{0B}
	- \partial_{l}f^{0A}\partial_{t}f_{0}^{B}) \bigr]
\end{multline*}
    and immediately see, that terms, proportional to
$ g^{0l} $
    contain combinations
\begin{align*}
    \frac{g^{0m}}{g^{00}} \partial_{l} f_{m}^{A} + \partial_{l} f_{0}^{A}
    =& \frac{1}{g^{00}} \partial_{l}f^{0A} -
    \frac{1}{g^{00}} (\partial_{l} g^{0m} f_{m}^{A} 
	+ \partial_{l}g^{00} f_{0}^{A}) \\
    \frac{g^{0m}}{g^{00}} \partial_{t} f_{m}^{A} + \partial_{t} f_{0}^{A}
	=& \frac{1}{g^{00}} \partial_{t} f^{0A} -
    \frac{1}{g^{00}} (\partial_{t} g^{0m} f_{m}^{A}
	+ \partial_{t}g^{00} f_{0}^{A})
\end{align*}
    and the second terms in the RHS are annihilated by vertical projector
$ \Pi^{AB} $.
    After this observation we see,
    that these terms cancel and we get the satisfactory expression
\begin{equation*}
    \Sigma = 2\sqrt{g} \Pi^{AB} \gamma^{kl} (
	\partial_{l}f_{k}^{A} \partial_{t} f^{0B}
	- \partial_{l}f^{0A} \partial_{t} f_{k}^{B}) .
\end{equation*}

    Let us do the same for
$ \HH $
    and separate the contributions, corresponding to
$ (\sigma,\rho)=(0,0)$, $(m,0)$,  $ (0,n) $ and $(m,n)$ .
    The 
$ (0,0) $
    component vanishes due to antisymmetry. The
$ (m,0) $
    and
$ (0,n) $
    components coincide after change of mute indeces and give
\begin{equation*}
    Q_{1} = 2\sqrt{g} \Pi^{AB} \gamma^{ln} g^{k0} (
	\partial_{k} f_{0}^{A} \partial_{l} f_{n}^{B}
	-\partial_{l} f_{0}^{A} \partial_{k} f_{n}^{B}) .
\end{equation*}
    The
$ (m,n) $
    contribution gives
\begin{equation*}
    Q_{2} = \sqrt{g} \Pi^{AB} g^{km} g^{ln} S_{kl,mn}^{AB} ,
\end{equation*}
    where
\begin{equation*}
    S_{kl,mn}^{AB} = \partial_{k}f_{m}^{A} \partial_{l}f_{n}^{B}
	- \partial_{k}f_{n}^{A} \partial_{l}f_{m}^{B} ,
\end{equation*}
    and substituting
$ g^{km} $ via
$ \gamma^{km} $, 
$ g^{0k} $,
$ g^{k0} $
    we get
$ Q_{2} = Q_{3}+Q_{4} $,
    where
\begin{equation*}
    Q_{3} = \sqrt{g} \Pi^{AB} \gamma^{km} \gamma^{ln} S_{kl,mn}^{AB}
\end{equation*}
    and
\begin{equation*}
    Q_{4} = 2\sqrt{g} \frac{g^{0k}}{g^{00}} \gamma^{ln} g^{0m}
	S_{kl,mn}^{AB} .
\end{equation*}
    Combining 
$ Q_{1} $ and
$ Q_{4} $
    and using the same trick as before we get
\begin{equation*}
    Q_{1}+Q_{4} = 2\sqrt{g} \frac{g^{0k}}{g^{00}} \Pi^{AB} \gamma^{ln} (
	\partial_{k} f^{0A} \partial_{l} f_{n}^{B}
	-\partial_{l} f^{0A} \partial_{k} f_{n}^{B} ) .
\end{equation*}
    Thus we get
\begin{gather*}
    \HH =  T_{0} + T_{1} \\
    T_{1} = Q_{1} + Q_{4} , \quad T_{0} = Q_{3} 
\end{gather*}
    and their expressions are satisfactory also.

    Now it is time to reduce the vertical projector
$ \Pi^{AB} $.
    We have
\begin{multline*}
    \Pi^{AB} = \delta^{AB} - g^{\mu\nu} f_{\mu}^{A} f_{\nu}^{B} = \\
	= \delta^{AB} - g^{ik} f_{i}^{A} f_{k}^{B} - 
    - g^{i0} f_{i}^{A} f_{0}^{B} - g^{0k} f_{0}^{A} f_{k}^{B}
      - g^{00} f_{0}^{A} f_{0}^{B} = \\
    = \delta^{AB} - \gamma^{ik} f_{i}^{A} f_{k}^{B} - \bigl(
	\frac{g^{0i}g^{0k}}{g^{00}} f_{i}^{A} f_{k}^{B}
	+ g^{0k} f_{0}^{A} f_{k}^{B} + \\
	+ g^{i0} f_{i}^{A} f_{0}^{B} + f_{0}^{A} f_{0}^{B} g^{00} \bigr) .
\end{multline*}
    The first two terms define the 3-dimensional vertical projector
    and the last can be rewritten as
\begin{equation*}
    \frac{1}{g^{00}} (g^{0i}f_{i}^{A} + g^{00}f_{0}^{A})
	(g^{0k}f_{k}^{B} + g^{00}f_{0}^{B}) = \frac{1}{g^{00}} f^{0A} f^{0B} .
\end{equation*}
    Thus we have
\begin{equation*}
    \Pi^{AB} = \delta^{AB} - \gamma^{ik} f_{i}^{A} f_{k}^{B} 
	- \frac{1}{g^{00}} f^{0A} f^{0B} .
\end{equation*}
    Let us mention, that the last term has proper normalization because
\begin{equation*}
    f^{0A} f^{0A} = g^{00} .
\end{equation*}
    Now I substitute this expression for
$ \Pi^{AB} $
    into
$ \Sigma $,
$ \HH_{0} $ and
$ \HH_{k} $.

    Begin with
$ \Sigma $:
    we get three contributions according to the form of
$ \Pi^{AB} $.
    The first is
\begin{equation*}
    \Sigma_{1} = 2 \sqrt{g} \gamma^{kl} (
	\partial_{l}f_{k}^{A} \partial_{t} f^{0A}
	- \partial_{l}f^{0A} \partial_{t} f_{k}^{A} ) .
\end{equation*}
    The last factor can be rewritten as
\begin{equation*}
    \partial_{t} (\partial_{l}f_{k}^{A}f^{0A})
    - \partial_{l} (\partial_{t}f_{k}^{A}f^{0A})
    = \partial_{t} \Omega_{kl}^{0} - \partial_{l} \Lambda_{k} ,
\end{equation*}
    where I remind the notation for
$ \Omega_{\alpha\mu}^{\beta} $
    and denote
\begin{equation*}
    \Lambda_{k} = \partial_{t} f_{k}^{A} f^{0A} .
\end{equation*}
    Thus we have
\begin{equation*}
    \Sigma_{1} = 2\sqrt{g} \gamma^{kl}(\partial_{t} \Omega_{kl}^{0}
	- \partial_{l} \Lambda_{k}).
\end{equation*}
    The first term here is quite satisfactory, it is almost of Darboux form.

    Now consider the second term
\begin{multline*}
    \Sigma_{2} = -2\sqrt{g} \gamma^{kl} \gamma^{mn} \bigl[
	(f_{m}^{A}\partial_{l}f_{k}^{A})(f_{n}^{B}\partial_{t}f^{0B})
	-(f_{m}^{A}\partial_{l}f^{0A})(f_{n}^{B}\partial_{t}f_{k}^{B})
	\bigr] = \\
    = 2\sqrt{g} \gamma^{kl} \gamma^{mn} \bigl( \omega_{m.kl} \Lambda_{n}
	- (f_{n}^{B}\partial_{t}f_{k}^{B}) \Omega_{ml}^{0} \bigr) .
\end{multline*}
    Here I used the orthonormality of
$ f_{k}^{A} $ and
$ f^{0A} $
    to rewrite
\begin{equation*}
    f_{n}^{B} \partial_{t} f^{0B} = -\partial_{t} f_{n}^{B} f^{0B} =
	- \Lambda_{n}
\end{equation*}
    and
\begin{equation*}
    f_{m}^{A} \partial_{l} f^{0A} = -\partial_{l} f_{m}^{A} f^{0A} =
	- \Omega_{ml}^{0}
\end{equation*}
    and introduce the 3-dimensional connection
\begin{equation*}
    \omega_{m,kl} = f_{m}^{A} \partial_{l} f_{k}^{A} .
\end{equation*}
    Finally the third contribution is given by
\begin{equation*}
    \Sigma_{3} = -\sqrt{g} \gamma^{kl} \frac{1}{g^{00}}
    [\Omega_{kl}^{0}\partial_{t}g^{00} -\partial_{l}g^{00}\Lambda_{k}] ,
\end{equation*}
    where I used that
\begin{equation*}
    f^{0A} \partial_{t} f^{0A} = \frac{1}{2} \partial_{t} g^{00} , \quad
    f^{0A} \partial_{l} f^{0A} = \frac{1}{2} \partial_{l} g^{00} .
\end{equation*}

    Let us collect all contrbutions containing
$ \Lambda_{k} $
\begin{equation*}
    \Lambda = 2\sqrt{g} \gamma^{kl} \bigl( -\partial_{l} \Lambda_{k}
	+ \omega_{kl}^{m} \Lambda_{m} + \frac{1}{2}
    \frac{\partial_{l}g^{00}}{g^{00}} \Lambda_{k} \bigr)
\end{equation*}
    and compare it with
\begin{multline*}
    \partial_{l} (\sqrt{g}\gamma^{kl}\Lambda_{k})
	= \partial_{l} \bigl(\frac{\sqrt{\gamma}}{\sqrt{g^{00}}}\gamma^{kl}
	    \Lambda_{k}\bigr) = \\
    = \sqrt{g} \bigl(\omega_{ml}^{m}
	- \frac{1}{2}\frac{\partial_{l}g^{00}}{g^{00}}\bigr)
	\gamma^{kl}\Lambda_{k} + \sqrt{g} \partial_{l}\gamma^{kl} \Lambda_{k}
	+ \sqrt{g} \gamma^{kl} \partial_{l}\Lambda_{k} .
\end{multline*}
    Using the vanishing of covarinat derivative of
$ \gamma^{kl} $
\begin{equation*}
    \partial_{l}\gamma^{kl} + \omega_{ml}^{k} \gamma^{ml}
	+ \omega_{ml}^{l} \gamma^{mk} = 0
\end{equation*}
    we get
\begin{equation*}
    \Lambda = - 2\partial_{l}(\sqrt{g}\gamma^{kl}\Lambda_{k})
	+2\sqrt{g}\gamma^{kl}\Lambda_{k} (\omega_{ml}^{m}-\omega_{lm}^{m}) .
\end{equation*}
    The second term in the RHS disappears due to mentioned above
    time-independent equations of motion. Indeed we have
\begin{multline*}
    \Omega_{ik}^{m} = f^{mA}\partial_{k}f_{i}^{A} = g^{m\sigma} f_{\sigma}^{A}
	\partial_{k}f_{i}^{A} = \\
   = \bigl(\gamma^{mn}+\frac{g^{m0}g^{n0}}{g^{00}}\bigr)
	f_{n}^{A} \partial_{k}f_{i}^{A} + g^{m0} f_{0}^{A}f_{i}^{A}
    = \omega_{ik}^{m} + \frac{g^{m0}}{g^{00}} \Omega_{ik}^{0}
\end{multline*}
    and 
$ \omega_{ik}^{m} $
    is symmetric to interchange
$ i \leftrightarrow k $
    together with
$ \Omega_{ik}^{m} $ and
$ \Omega_{ik}^{0} $.

    Thus the full contribution containing
$ \Lambda_{k} $
    is a pure divergence and can be omitted.

    Consider now the expression
\begin{equation*}
    \gamma^{kl} \gamma^{mn} \Omega_{ml}^{0} (f_{n}^{A} \partial_{t}f_{k}^{A}).
\end{equation*}
    Due to symmetry of
$ \Omega_{ml}^{0} $
    it can be rewritten as
\begin{equation*}
    \frac{1}{2} \gamma^{kl}\gamma^{mn} (f_{n}^{A}\partial_{t}f_{k}^{A} +
	f_{k}^{A}\partial_{t}f_{n}^{A}) \Omega_{ml}^{0}
    = - \frac{1}{2} \partial_{t} \gamma^{ml} \Omega_{ml}^{0} .
\end{equation*}
    Using this we have
\begin{multline*}
    \Sigma = \sqrt{g} \bigl( 2\gamma^{kl}\partial_{t}\Omega_{kl}
	+\partial_{t}\gamma^{kl} \Omega_{kl}^{0} - \gamma^{kl} \Omega_{kl}
	\frac{\partial_{t}g^{00}}{g^{00}} \bigr) = \\
    = q^{kl}\partial_{t}\Pi_{kl}
	+ \partial_{t} (\sqrt{g}\gamma^{kl}\Omega_{kl}) ,
\end{multline*}
    where
\begin{equation*}
    q^{kl} = \gamma \gamma^{kl} , \quad
	\Pi_{kl} = \frac{1}{\sqrt{\gamma g^{00}}} \Omega_{kl}^{0} .
\end{equation*}
    The second term can be dropped and we obtain the canonical expression for
    the one form
$ \Sigma $.
    The normalization of canonical pairs --- 
$ q^{ik} $
    as contravariant density of weight 1 and
$ \Pi_{ik} $
    as covariant density of weight
$ -1/2 $ ---
    appeared first in the paper of Schwinger
\cite{Schwinger}.
    I used it in the survey
\cite{survey}.

    Let us turn now to 
$ \HH $. First take
$ T_{1} $
\begin{equation*}
    T_{1} = 2\sqrt{g}\frac{g^{k0}}{g^{00}} \Pi^{AB} \gamma^{lm}
    (\partial_{k}f^{0A}\partial_{l}f_{m}^{B} 
	-\partial_{l}f^{0A}\partial_{k}f_{m}^{B})
\end{equation*}
    and consider three contrbutions according to form of
$ \Pi^{AB} $.

    In the first we use
\begin{multline*}
    \partial_{k}f^{0A}\partial_{l}f_{m}^{A}
    - \partial_{l}f^{0A}\partial_{k}f_{m}^{A} 
    = \partial_{k} (f^{0A} \partial_{l}f_{m}^{A})
	- \partial_{l} (f^{0A} \partial_{k}f_{m}^{A}) = \\
    = \partial_{k} \Omega_{ml}^{0} - \partial_{l} \Omega_{mk}^{0} .
\end{multline*}
    In the second we get
\begin{multline*}
    -\gamma^{pq} \bigl[
	(f_{p}^{A}\partial_{k}f^{0A}) (f_{q}^{B}\partial_{l}f_{m}^{B})
	-(f_{p}^{A}\partial_{l}f^{0A}) (f_{q}^{B}\partial_{k}f_{m}^{B})
    \bigr] = \\
    = \omega_{ml}^{p} \Omega_{pk}^{0} - \omega_{mk}^{p} \Omega_{pl}^{0} .
\end{multline*}
    Finally in the third term we have
\begin{multline*}
    -\frac{1}{g^{00}} \bigl[
	(f^{0A}\partial_{k}f^{0A}) (f^{0B}\partial_{l}f_{m}^{B})
	-(f^{0A}\partial_{l}f^{0A}) (f^{0B}\partial_{k}f_{m}^{B})
    \bigr] = \\
    = -\frac{1}{2} \frac{\partial_{k}g^{00}}{g^{00}} \Omega_{ml}^{0}
	+ \frac{1}{2} \frac{\partial_{l}g^{00}}{g^{00}} \Omega_{mk}^{0} .
\end{multline*}
    Combining all together we get
\begin{multline*}
    T_{1} = 2 \frac{g^{k0}}{g^{00}} \sqrt{g} \gamma^{lm} \bigl[
	\partial_{k} \Omega_{ml}^{0} - \partial_{l}\Omega_{mk}^{0}
    -\frac{1}{2}\frac{\partial_{k}g^{00}}{g^{00}} \Omega_{ml}^{0} + \\
    + \frac{1}{2} \frac{\partial_{l}g^{00}}{g^{00}} \Omega_{mk}^{0}
	+\omega_{ml}^{p} \Omega_{pk}^{0} + \omega_{mn}^{p}\Omega_{pl}^{0}
	\bigr] .
\end{multline*}
    Taking into account symmetry of
$ \omega_{ml}^{p} $
    and 
$ \Omega_{pl}^{0} $
    this can be written as
\begin{equation*}
    T_{1} = \lambda^{k} \HH_{k} ,
\end{equation*}
    where we introduce Lagrange multipliers
\begin{equation*}
    \lambda^{k} = \frac{g^{0k}}{g^{00}}
\end{equation*}
    and
\begin{equation*}
    \HH_{k} = 2 \bigl[
	\nabla_{k}(q^{ml}\Pi_{ml}) -\nabla_{l}(q^{ml}\Pi_{mk}) \bigr] .
\end{equation*}

    Finally consider
$ T_{3} $
    and take into account, that the first two terms in
$ \Pi^{AB} $
    give 3-dimensional analogue of the vertical projector. Its contribution to
$ T_{0} $
    is
\begin{equation*}
    \sqrt{g} \gamma^{km} \gamma^{ln} \Pi^{AB(3)} S_{kl,mn}^{AB}
	= \sqrt{g} S^{(3)} ,
\end{equation*}
    where
$ S^{(3)} $
    is scalar curvature of metric
$ \gamma_{ik} $
    and connection
$ \omega_{nl}^{m} $.
    The last term in 
$ \Pi^{AB} $
    gives
\begin{equation*}
    -\frac{f^{0A}f^{0B}}{g^{00}} S_{kl,mn}^{AB} = -\frac{1}{g^{00}}
	(\Omega_{mk}^{0}\Omega_{nl}^{0} -\Omega_{nk}^{0}\Omega_{ml}^{0}) .
\end{equation*}
    With this
$ T_{0} $
    can be rewritten as
\begin{equation*}
    T_{0} = \lambda_{0} \HH_{0} ,
\end{equation*}
    where Lagrange multiplier 
$ \lambda_{0} $
    is given by
\begin{equation*}
    \lambda_{0} = \frac{\sqrt{g}}{\gamma} = \frac{1}{\sqrt{g^{00}}\gamma}
	= \frac{1}{\sqrt{g}g^{00}}
\end{equation*}
    and
\begin{equation*}
    \HH_{0} = \gamma S^{(3)} - q^{km}q^{ln}
	(\Pi_{km}\Pi_{ln} - \Pi_{lm}\Pi_{kn}) .
\end{equation*}

    This finishes calculations in this note. Let us remind that in it we used
    the change of 40 variables
$ f_{\mu}^{A} $
    to the set
$ (f_{k}^{A}, f^{0A}, g^{00}, g^{0k}) $.
    Superficially we have here 44 components, however we have 4 constraints
\begin{equation*}
    f_{k}^{A} f^{0A} = 0 , \quad f^{0A} f^{0A} = g^{00} .
\end{equation*}
    The main result is the formula for the action in 3+1 decomposition
\begin{equation*}
    A = \int d^{3}x \int dt (q^{ik}\partial_{t}\Pi_{ik} + \lambda^{0}\HH_{0}
	+\lambda^{k}\HH_{k}) ,
\end{equation*}
    which coincides with formulas of Dirac and ADM.
    Thus it is shown once more, that my proposal is equivalent to
    the classical formalism of Hilbert-Einstein.

    However, as I already said in the beginning, this formulation could
    be a point of departure for the generalization not evident in the
    classical formulation.

\end{document}